\documentclass[12pt]{naturex}
\pdfoutput=1
\usepackage[T1]{fontenc}
\RequirePackage{cite}
\usepackage{amsmath}
\usepackage{times}
\usepackage{geometry}                
\usepackage{float}
\usepackage{subfloat}
\usepackage{subfig}
\usepackage{caption}
\usepackage{graphicx}
\usepackage{amssymb}
\usepackage{epstopdf}
\usepackage{wrapfig}
\usepackage{floatpag}

\def\blfootnote{\xdef\@thefnmark{}\@footnotetext}

\long\def\symbolfootnote[#1]#2{\begingroup%
\def\thefootnote{\fnsymbol{footnote}}\footnote[#1]{#2}\endgroup}

\long\def\nullfootnote[#1]#2{\begingroup%
\def\thefootnote{#1}\footnote[#1]{#2}\endgroup}

\newfloat{boxfigure}{thp}{lop}
\floatname{boxfigure}{Box Figure}
\newsubfloat{boxfigure}

\long\def\symbolfootnote[#1]#2{\begingroup%
\def\thefootnote{\fnsymbol{footnote}}\footnote[#1]{#2}\endgroup}

\title{Quantifying the interdisciplinarity of scientific journals and fields.}
\author{Filipi N. Silva\nullfootnote[1]{Institute of Physics of S\~ao Carlos, University of S\~ao Paulo.$^\dagger$Present address: PO Box 369, S\~ao Carlos, S\~ao Paulo, 13560-970, phone +55 16 3373 9858, Brazil (ldfcosta@gmail.com).}, Francisco A. Rodrigues\nullfootnote[2]{Institute of Mathematics and Computer Sciences, University of S\~ao Paulo.}, Osvaldo N. Oliveira Jr$^1$, Luciano da F. Costa$^{1,}$\nullfootnote[3]{National Institute of Science and Technology for Complex Systems, Rio de Janeiro, Brazil}~~$^\dagger$}
\date{}
\begin{document}

\maketitle 

\begin{abstract}
There is an overall perception of increased interdisciplinarity in science, but this is difficult to confirm quantitatively owing to the lack of adequate methods to evaluate subjective phenomena. This is no different from the difficulties in establishing quantitative relationships in human and social sciences\cite{eto:2003}. In this paper we quantified the interdisciplinarity\cite{raan:2000p6678} of scientific journals and science fields by using an entropy measurement\cite{Duda:2001qc,Lee:2010zr} based on the diversity of the subject categories of journals citing a specific journal. The methodology consisted in building citation networks using the Journal Citation Reports database, in which the nodes were journals and edges were established based on citations among journals. The overall network for the 11-year period (1999-2009) studied was small-world\cite{Watts:1998p1593} and scale free\cite{Albert:2002p161} with regard to the in-strength. Upon visualizing the network topology an overall structure of the various science fields could be inferred, especially their interconnections. We confirmed quantitatively that science fields are becoming increasingly interdisciplinary, with the degree of interdisplinarity (i.e. entropy) correlating strongly with the in-strength of journals and with the impact factor.
\end{abstract}
\clearpage

The tremendous advances in scientific methods and increase in computational power in the last few decades have allowed increasingly complex problems to be addressed and solved\cite{Barabasi:2004uq,Costa:2011fk}. Because such types of problems are intrinsically interdisciplinary, this has reinforced the pan-multidisciplinary nature of many naturally-occurring phenomena and man-made systems. In a sense, this movement brought science closer to the paradigm adopted by Greek philosophers who treated Nature as a landscape of knowledge glued together in an indivisible discipline.  Not surprisingly, in recent years new areas have been established with this interdisciplinary character, as is the case of nanoscience and nanotechnology, in addition to new disciplines arising from the merging of two or more areas, such as computational biology and biomolecular physics. The interdisciplinary global structure of knowledge has not received much attention in the literature, probably due to the difficulty in quantifying how interdisciplinary a given topic or piece of work is \cite{eto:2003,raan:2000p6678,Costa:2008p14}. A possible approach to deal with such intricate relationships is to treat large systems as complex networks\cite{Albert:2002p161,Newman:2003p274}, which are convenient to represent complex system structures where subsystems are the vertices and their interactions are represented by edges in a graph. Though built from simple elements, these networks may present high complexity both in size and in topology\cite{Barabasi:1999p279}, thus providing an adequate framework to capture the complex behavior of systems without narrowing the study to simple, isolated systems.

In this paper we used concepts from complex networks to evaluate quantitatively the interdisciplinarity of science fields and journals. The citation networks were built in a different manner from the conventional one employed in the literature. Rather than taking a paper (or any item in the literature) as a node\cite{Newman:2001p264,Newman:2004p277,Newman:2001p270,Newman:2001p269,Redner:1998p2563}, we built the network with journals, indexed in the \emph{Journal Citation Reports}\symbolfootnote[2]{{http://thomsonreuters.com/products\_services/science\_products/a-z/journal\_citation\_reports/} \\ URL Retrieved on Feburary 28, 2011.} (JCR) database, being the nodes and the links being established from citations between journals. The main reason for this choice is that the network generated can be handled computationally, which otherwise would be difficult to do for the large size of conventional citation networks. Furthermore, because the JCR database is not a subgraph of a larger structure, it may provide a better overview of the structure of knowledge than using arbitrary subnetworks of articles citation networks. This has been done by obtaining metrics of the topology of the journal citation network and assessing the interdisciplinarity of a journal or a field by analyzing the diversity of the nodes linked to a specific journal.

\begin{figure*}[!h]
  \centering
  \includegraphics[width=10cm]{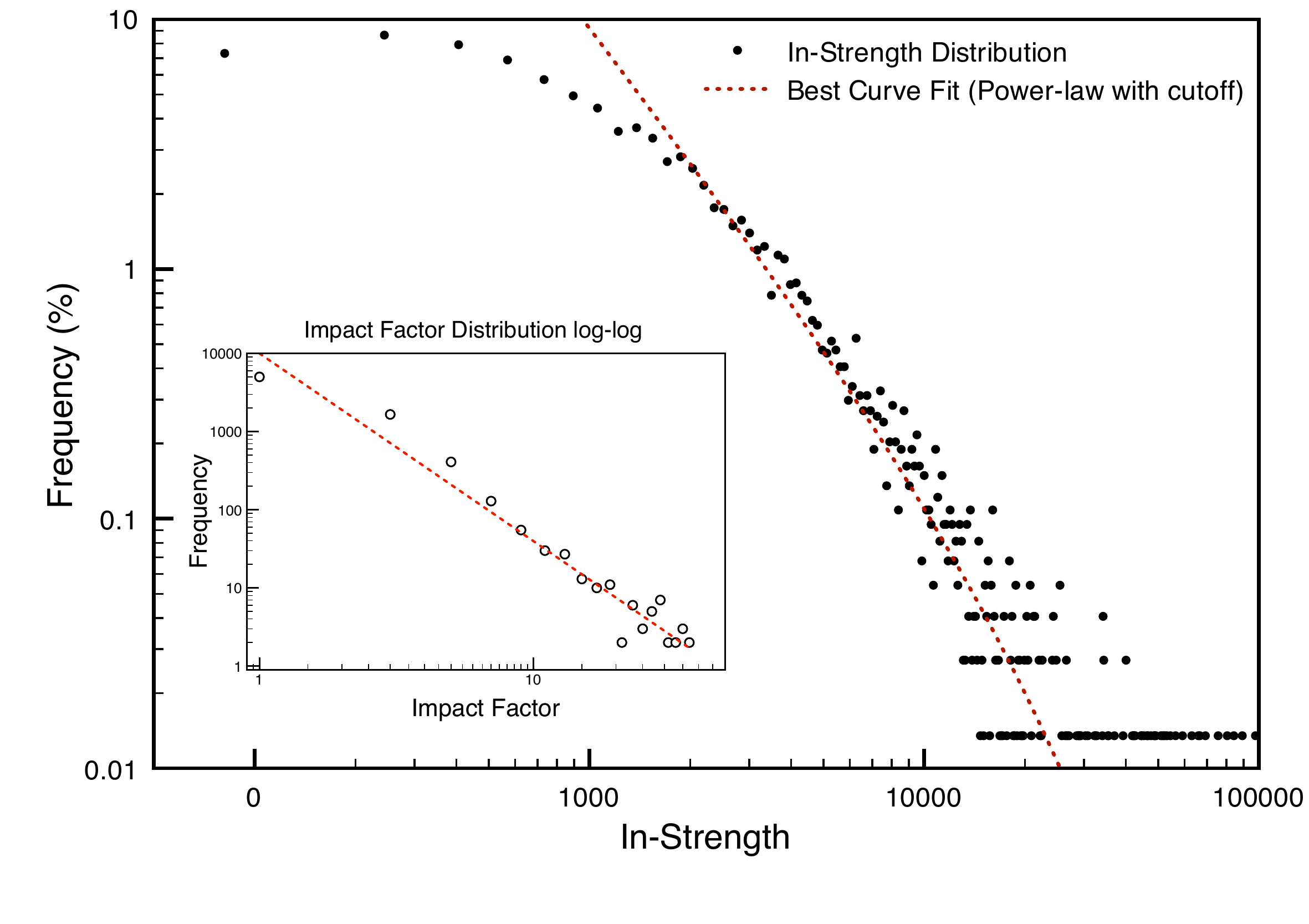}
 \caption{{\bf In-Strength and Impact Factor Distributions.} The in-strength distribution resulted in a power law with exponential cuttoff\cite{Clauset:2009vn}, where the best fitting $F(k)=k^\alpha e^{\beta k}$ is shown as a dashed red line with power coefficient $\alpha=-1.73$. The inset shows that the impact factor distribution also obeys a power law best fitted by the curve shown in red with a power coefficient $\alpha=-2.4$. }
  \label{fig:distributions}
\end{figure*}

The network was built with the nodes representing all the $7387$ journals indexed in the JCR database and the edges were established considering citations during the $1999$-$2009$ period. The edges were directed and weighted, with the weight being the number of citations from one journal to the other. The subject categories and the major science fields assigned to each journal (described in the methodology) were also extracted from the JCR database. The resulting network is scale-free\cite{Albert:2002p161} in terms of the in-strength with a power-law distribution with cutoff\cite{Clauset:2009vn}, as shown in Figure~\ref{fig:distributions}. It is also small-world\cite{Watts:1998p1593}, since its average shortest path is $2.4$ and the maximum shortest path (or network diameter) is~$5$.

\begin{figure*}[!h]
	\includegraphics[width=14.0cm]{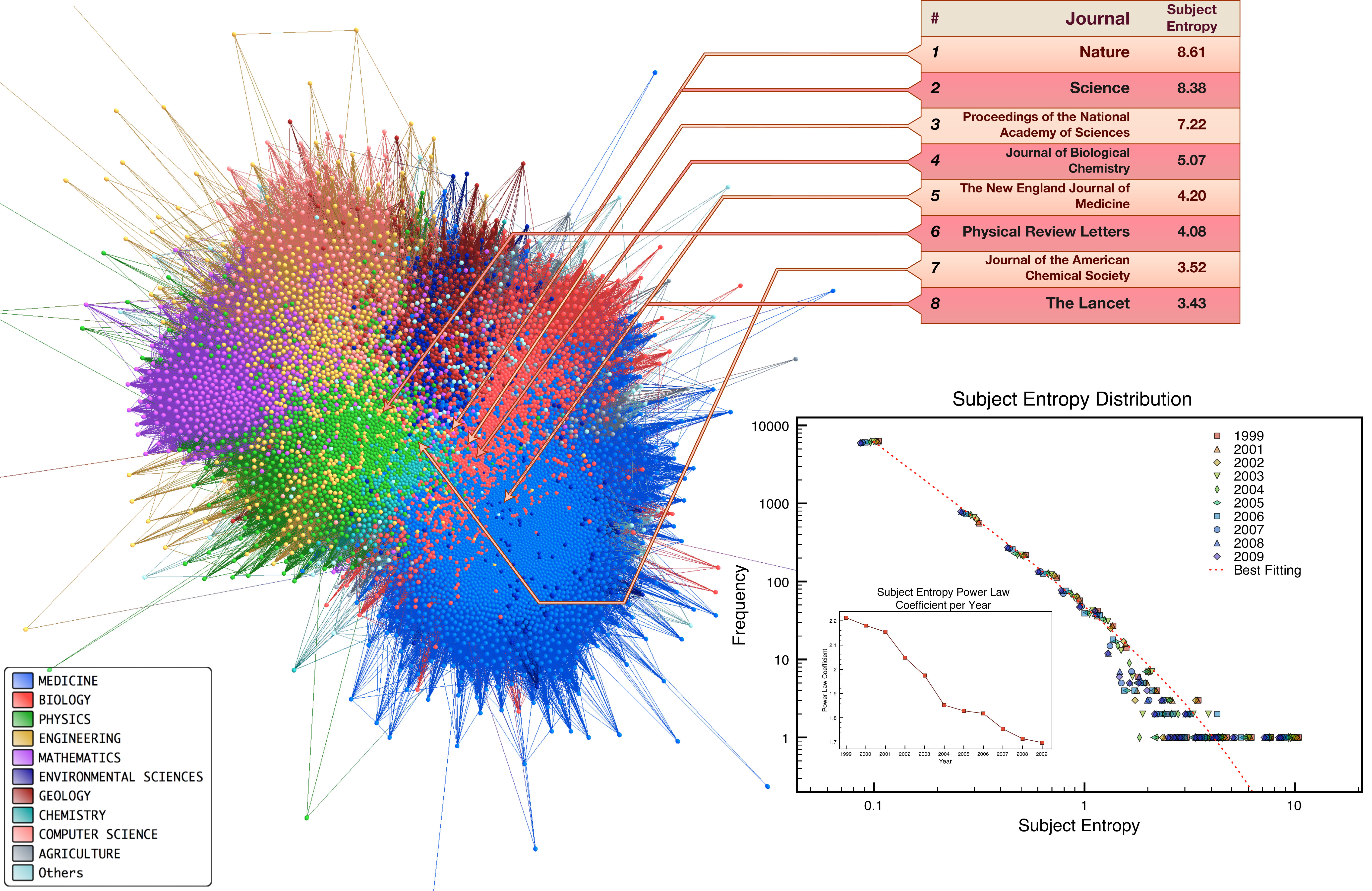}
  \caption{{\bf Subject Entropy and its relation with subject categories.} The main figure shows the planar projection of the network for the year 2009 with colors representing subject categories according to the color legend. The table on the top right shows the 8 journals with highest entropy. The distribution of subject category in the bottom-right panel obeys a power law with cuttoff with the best fitting in the dashed red curve for a coefficient $\alpha=1.97$. The insert in the panel indicates a decreasing coefficient for the power law as time goes by. This means an increased diversity of the values of subject entropy. In obtaining these results, the self-citations among journals were not eliminated. Nevertheless, in subsidiary experiments we found that the exclusion of self-citations has little effect on the overall properties of the network.}
  \label{fig:networkProjection}
\end{figure*}

The metrics\cite{Costa:2007kx} \emph{node in-strength} and \emph{betweenness centrality}\cite{Brandes:2001p2678} were obtained for each subnetwork defined by the subject categories. The formal definition of the metrics is given in the Methodology. The network nodes were projected onto a 2D space using force-directed methods\cite{KAMADA:1989lr,FRUCHTERMAN:1991fk,Greengard:1987p908} (see Methodology for details), which display the interconnection between subject categories and science fields, as shown in figure~\ref{fig:networkProjection}. This mapping can be understood as a low dimensional representation of the network\cite{Davidson:1996uq,Cohen:1997kx,Chen:2009ly} and provides information on the topological proximity between journals in the network. Medical disciplines appear together, alongside veterinary disciplines while mathematics (pure) appears isolated, being connected to the giant component by engineering and applied mathematics. Biology and Molecular biology create the link between biological/medical sciences and exact sciences. Geosciences, Plant Sciences and Environmental Sciences form a very compact group. Medicine is the most representative group in the network, with the highest number of journals (ca. 34\% of the journals).

\begin{figure*}[ph!]
\floatpagestyle{empty}
	\includegraphics[width=10.0cm]{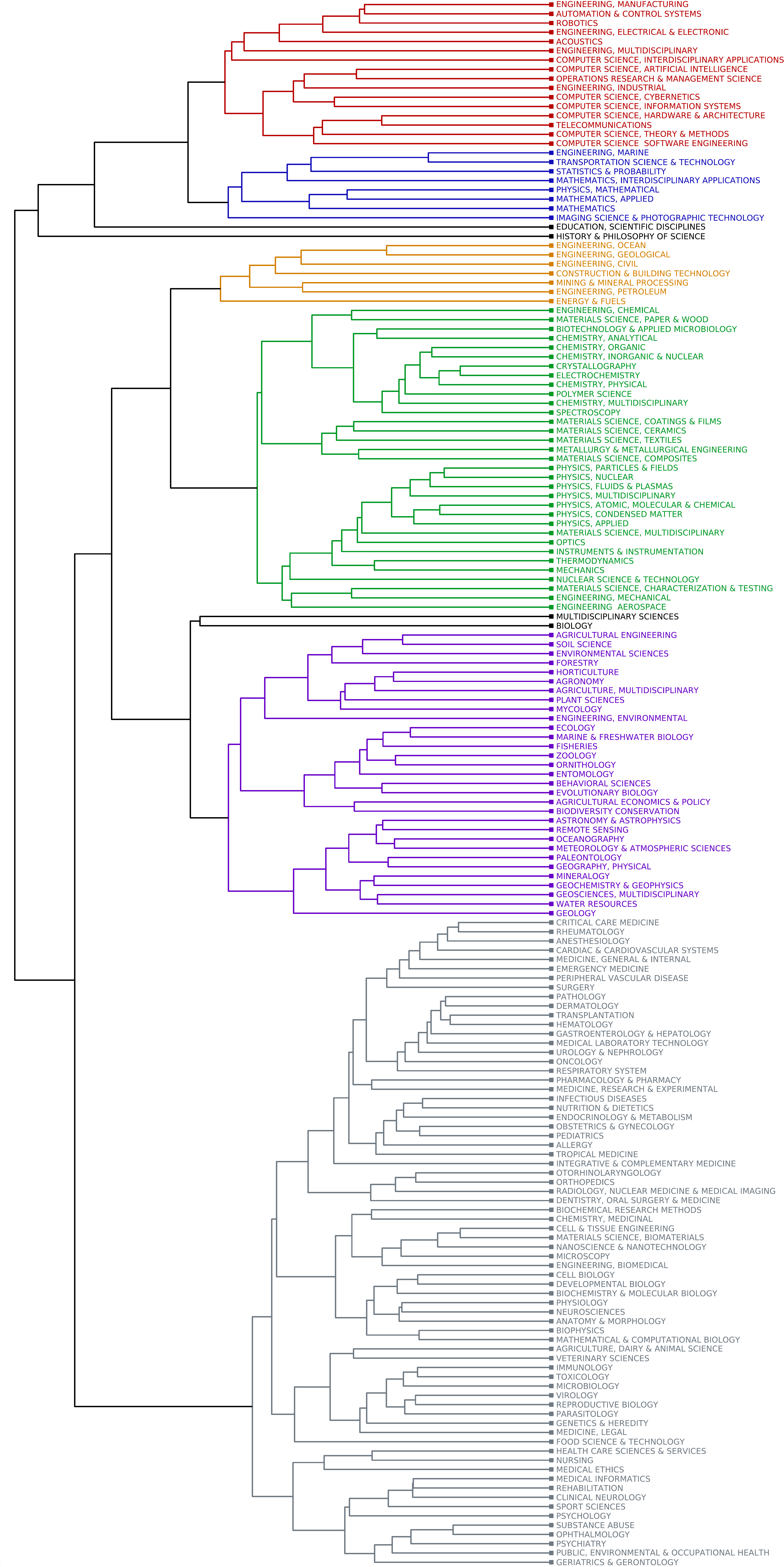}
  \caption{{\bf Dendrogram of subject categories} . Each subject category is presented in the dendrogram with colors corresponding to every grouped  journals  considering a cut on the dendrogram about the middle of the distances.}
  \label{fig:dendrogram}
\end{figure*}

Another approach to visualize the structure of knowledge is to obtain a dendrogram representing the projection of the network topology\cite{Costa:2009ss,Duda:2001qc}. The dendrogram was obtained by agglomerative hierarchical clustering and considering the average linkage and the topological distance (average of shortest paths).  Figure~\ref{fig:dendrogram} shows the dendrogram with different colors for distinct fields. An inspection of the dendrogram confirms what was inferred from the 2D projections. For instance, Mathematics and Computer Sciences are close together, as one should expect. Engineering is connected with Mathematics and Computer Science. Physics and Chemistry are very close, with Chemistry making the connection between Physics and Biological Sciences, while Biology connects Medicine and Exact Sciences.

\begin{figure*}[!h]
  \centering
  \subfloat[]{\label{fig:averageShortestPathPerYear}
  \includegraphics[width =5cm]{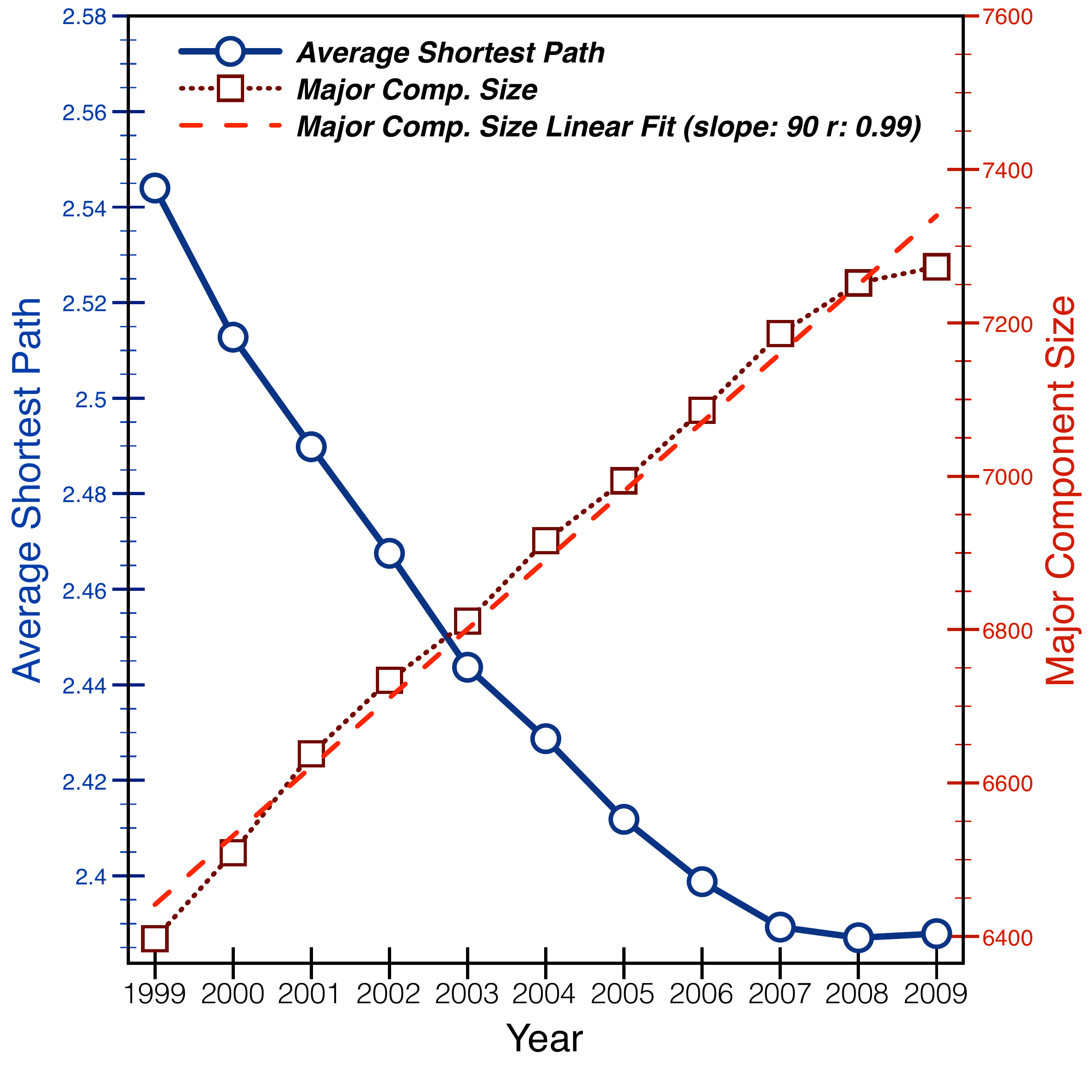}
  }
~
  \subfloat[]{\label{fig:averageIn-DegreePerGroup}
  \includegraphics[width=5cm]{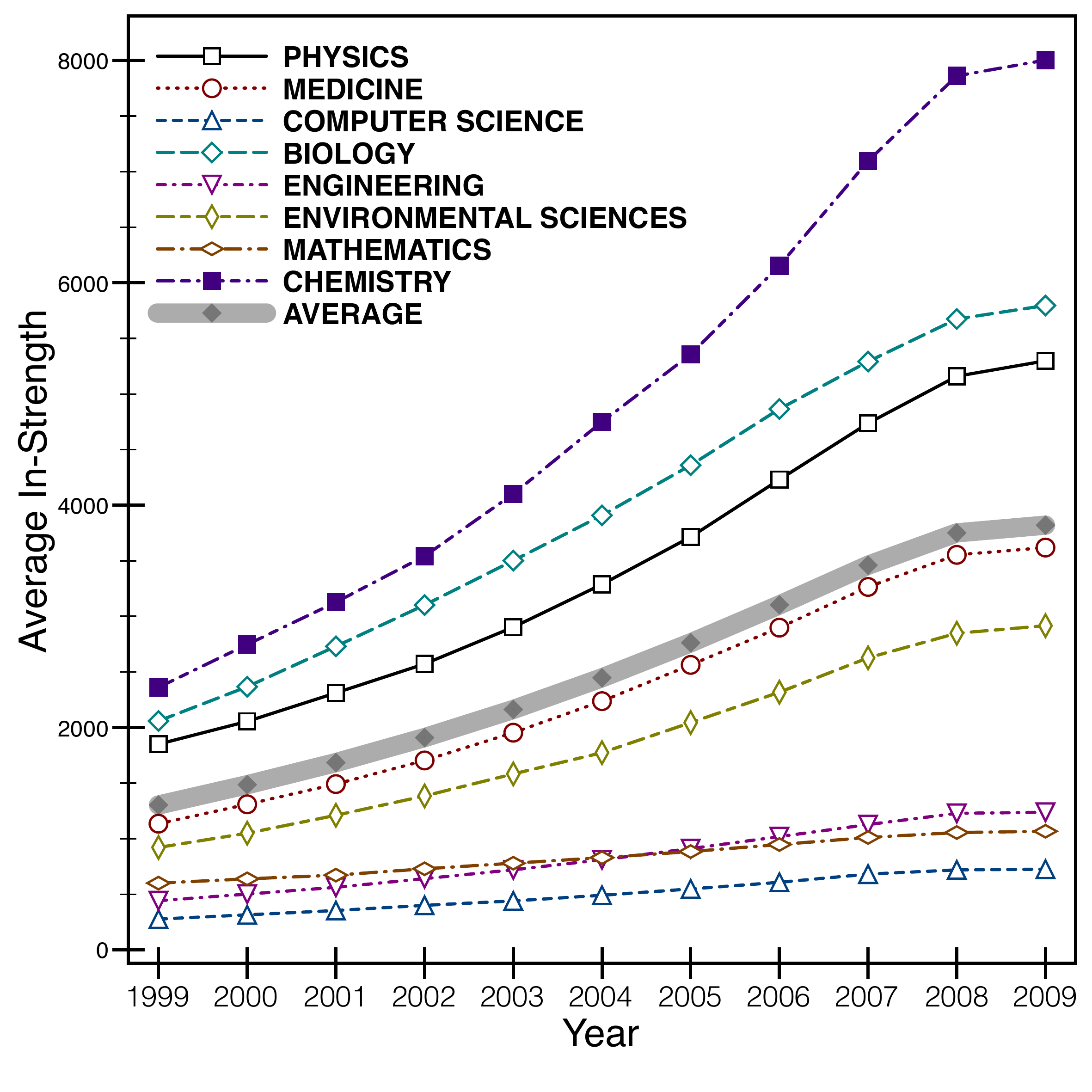}
  }
~
  \subfloat[]{\label{fig:medianSubjectEntropyPerGroup}
  \includegraphics[width=5cm]{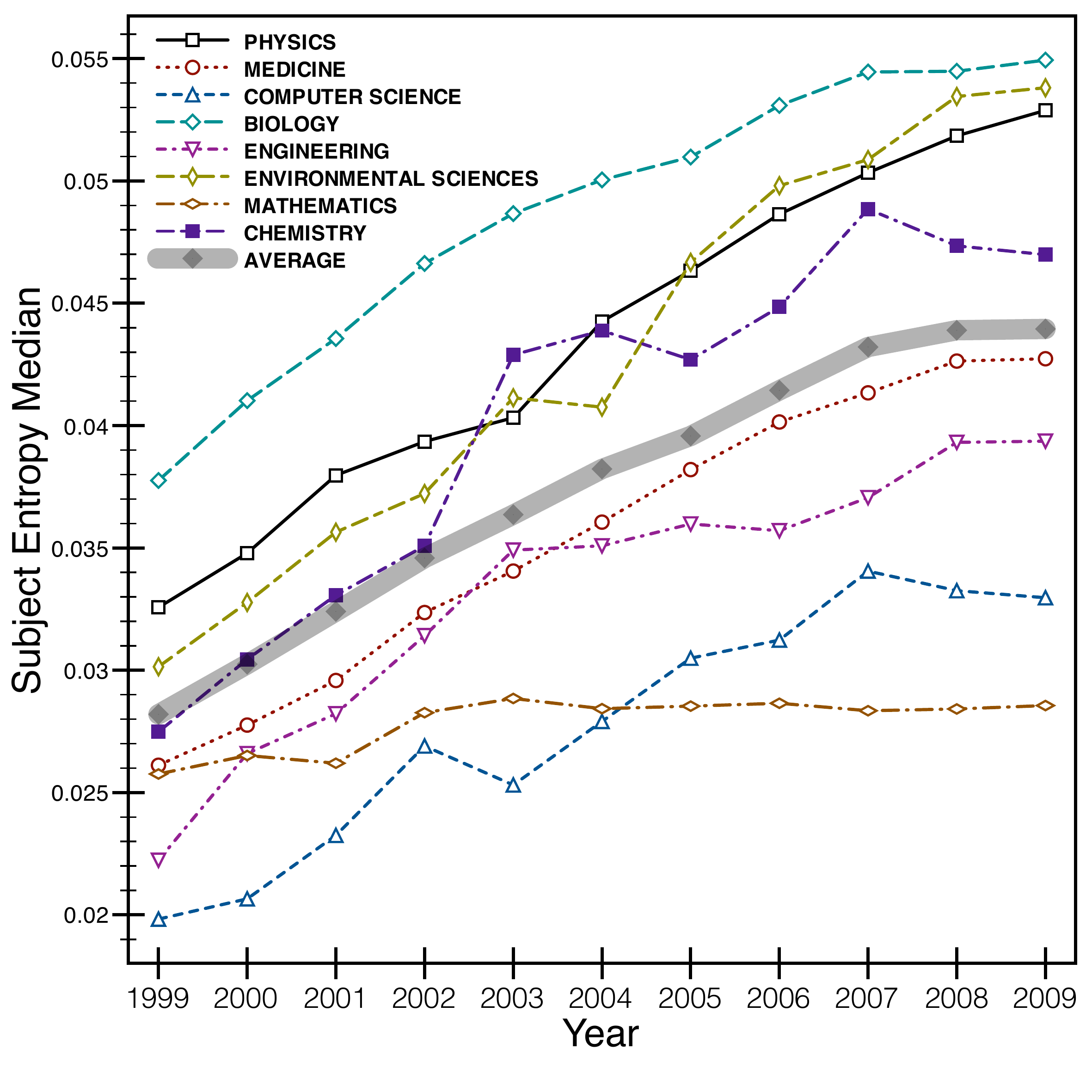}
  }
  \caption{ {\bf Time evolution of the average or median of the various measurements.} {\bf a} The size of the major component of the citation network is depicted in the red curve, displaying an almost linear increase with time $(r=0.99)$ indicated by the dashed red line. The average shortest path shown in blue decreases monotonically with time.  {\bf b}, {\bf c} Average in-strength and median of the subject entropy versus time for several major science fields according to the subject categories (see Methodology). The global values, \emph{i.e.} considering all journals, are represented in both panels by a thicker curve in gray.
  }
  \label{fig:perYearMeasurements}
\end{figure*}

	The temporal evolution of the journal citation network is depicted in figure~\ref{fig:perYearMeasurements}. The size of the main component of the network increased with time, as one should expect from the increase in the number of journals. Indeed, the average shortest path decreased with time until 2006, as shown in figure~\ref{fig:averageShortestPathPerYear}. This stabilization may be ascribed to a quasi saturation in the network growth, also shown in the figure. The in-strength of any given node (i.e. journal) correlates with its impact, for it is given by the total number of citations received by the journal. The impact of some areas has increased considerably in recent years, as illustrated in figure~\ref{fig:averageIn-DegreePerGroup}. This is the case of chemistry, biology and physics, whose in-strengths were already high. The temporal evolution of the whole network (average in the figure) almost coincides with that of medicine, probably because the medicine journals comprise $34\%$ of the whole network. Interestingly, medicine is not the most cited field, which is reflected in a poor correlation between the number of articles and the in-strength of the journals. As we shall show later on, the higher impact correlates well with the interdisciplinary nature of the field. 
	
	Also shown in figure~\ref{fig:medianSubjectEntropyPerGroup} is that over the years the fields have become more interdisciplinary, thus confirming the overall perception mentioned before. The interdisciplinarity index was introduced to measure the diversity of subject categories for the citation neighborhood of a journal. It is defined as the Shannon entropy of the subject categories histograms obtained from the immediate neighborhood for each journal (see the formal definition in the Methodology). Therefore, the higher the entropy the more interdisciplinary a journal is. The same applies to fields, as the data were collected from the journals representing a specific field. The average entropy for the main fields varied with time according to figure~\ref{fig:medianSubjectEntropyPerGroup}. The impact of a field -- as quantified in terms of citations its journals receive -- tends to increase with the interdisciplinary nature. Indeed, Table~\ref{table:correlations} shows a very high correlation between the in-strength and the Shannon entropy for all journals. Most significantly, the highest correlation for the impact factor occurred for the subject entropy. Particularly high entropies were obtained for journals with very wide readership, which publish work in any field of science as is the case of the three highest entropies. These journals are followed by those from specific fields, but that again have a wide readership, as indicated in the table accompanying figure~\ref{fig:networkProjection}. The other network metric with high correlation with the subject entropy was the betweenness centrality, which is normally a measure of importance of nodes in a network. 
		
\begin{table*}[!htbp]
  \centering
  \includegraphics[width=12cm]{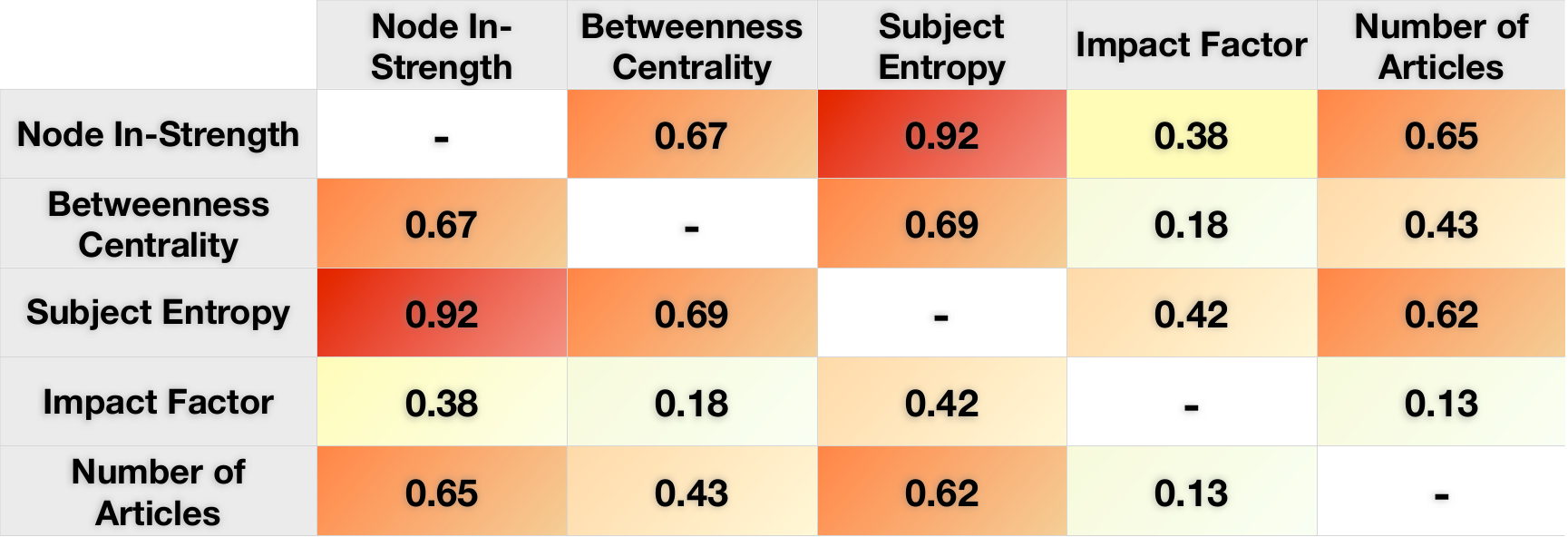}
 \caption{{\bf Correlation between the network metrics and other features of the citation network}. Some of the correlations are intuitive, such as those associated with the number of papers, which correlates highly with the in-strength and subject entropy but poorly with the impact factor. In other words, both the subject entropy and the in-strength should scale with the size of the journal in terms of number of papers. Another expected correlation appeared between the subject entropy and the in-strength, for the latter reflects the number of citations. The impact factor has the highest correlation with the subject entropy, thus indicating that increasing interdisciplinarity causes an increase in impact.}
  \label{table:correlations}
\end{table*}

	The distribution of journals according to their entropies also obeys a power law, as shown in Figure~\ref{fig:networkProjection}, which means that the majority of journals are dedicated to specific topics, as one should expect. This was observed for networks considering or disregarding self-citations. The insert in Figure~\ref{fig:distributions} shows that a power law also applies to the distribution of journals according to their impact factors. 

In summary, the combination of a new measure for interdisciplinarity exploiting the subject entropy and a novel way to build a citation network allowed us to identify the most interdisciplinary fields and their interconnections. Chemistry, Physics and Biology have been found highly interdisciplinary, as expected, but surprisingly there is relatively little interdisciplinarity in computer science (though it has increased recently). The visualization of the citation network also served to illustrate relationships between distinct science fields. With the generality of the approaches proposed here, the way is paved for ontologies for science and technology to be constructed, in addition to providing important information for research and development policy makers.

\begin{addendum}
 \item This work was supported by FAPESP, CAPES, and CNPq (Brazil).
 \item[Correspondence] Correspondence and requests for materials
should be addressed to Luciano da Fontoura Costa~(email: ldfcosta@gmail.com).
 \end{addendum}
 
\clearpage
\bibliographystyle{naturemag}
\bibliography{JournalsArticle}

\clearpage
\newpage
\begin{center}
{\large Methodology}
\end{center}
\subsection{Journal Citation Networks}
The complete set of indexed scientific journals was obtained in an automated fashion from the database of \emph{Journal Citation Reports} (JCR). Other pieces of information collected were the impact factor, subject categories and citations per paper for the 11-year period between $1999$ and $2009$. A complex network was obtained for the whole period, and for each year separately, mapping journals as nodes and citations from pair of journals as edges in such a way that the networks grow incrementally over time. For example, the network corresponding to the year $2005$ contains the network of $2004$ as well the one from $2003$ and so on.

\begin{figure*}[!h]
  \centering
  \subfloat[]{\label{fig:journalNetworkMethod}
  \includegraphics[width=7.25cm]{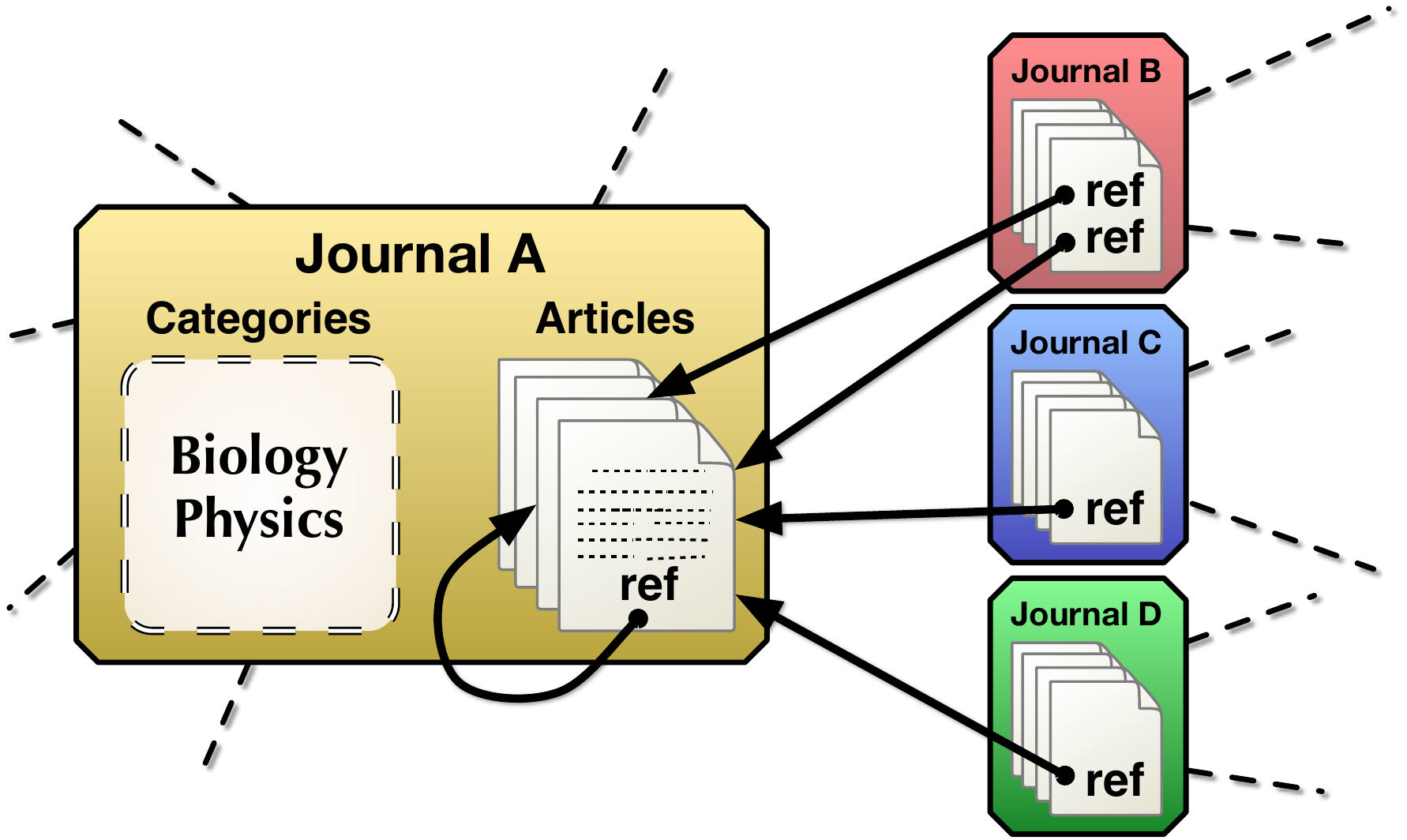}
  }
~
  \subfloat[]{\label{fig:kwnoledgeAreasNetwork}
  \includegraphics[width =7.25cm]{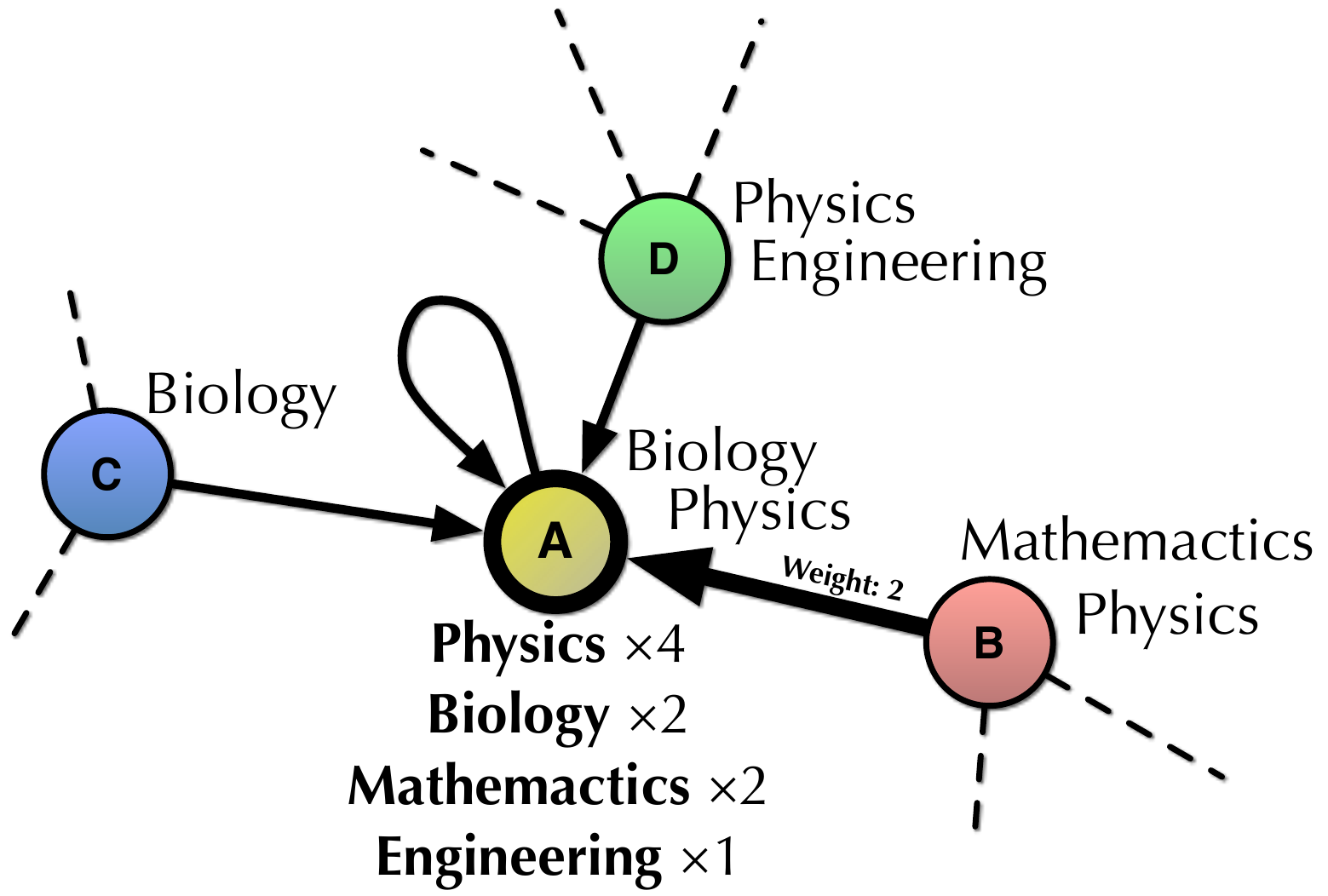}
  }
  \caption{{\bf Schematic representation of the procedure to build the network and its structure.}  In {\bf a} journal $A$ was connected to the network by identifying citations from each article from $A$ to any journal in the network, including itself. In the example, $A$ is connected to journals $B$, $C$ and $D$. A set of subject categories, $\{Biology, Physics\}$, is also associated with Journal A. {\bf b} depicts the interconnection between journals $A$, $B$ and $C$, and related edges weights, as well the frequency histogram of subject categories for the neighborhood of journal $A$, shown as the total count of appearances after their names.}
  \label{fig:methodology}
\end{figure*}

Because of the nature of the journal citation structure, the networks allow self-loops and are directed. Also, they are edge weighted so that the strength of a connection is directly related to the number of citations between papers from a pair of journals. Figure~\ref{fig:journalNetworkMethod} depicts the structure of these networks along with the subject categories.

\subsection{Entropy as a Measurement of Interdisciplinarity}
The interdisciplinarity of a journal can be understood as being related to how diverse, in terms of their subject categories, the journals citing the journal under analysis are. It is similar to the \emph{Shannon disparity}\cite{Lee:2010zr}, which quantifies the heterogeneity of the weights of edges coming from a reference node by using information entropy considering edge weights histograms. A similar measurement -- now related to the heterogeneity of subject categories of the neighborhood of a node -- can be obtained with the entropy $H_j$ of probabilities $P_j(c)$, for each subject category $c$ presented on the citing neighborhood of a journal. The JCR database provides a set of $172$ subject categories in a way that each journal is coupled with a subset of at least one subject category.  

In order to balance the strength of each category, the probability $P_j(c)$ was taken into account as the normalized sum of probabilities of a journal $j$, having a category $c$, i.e. $P_j(c)$ was obtained by normalizing each subject category histogram, $h_j(c)$, by the total frequency of each category considering all journals, as given by equation \ref{eq:probability}: 

\begin{align}
\label{eq:probability}
\bar{h}_j(c) &= {{h_j (c) \over \sum_{j} h_j (c)}} \\
P_j(c) &= {\bar{h}_j(c) \over \sum_{c} \bar{h}_j(c))}
\end{align}

The measurement of interdisciplinarity can be obtained by simply taking the classical information entropy of the proposed normalized probabilities of subject categories, as given by equation~\ref{eq:entropy}.

\begin{equation}
\label{eq:entropy}
H_j = -\sum_{\text{c}} \begin{cases} P_j(c) \ln (P_j(c)) & \mbox{if } P_j(c) \neq 0 \\ 0 & \mbox{if } P_j(c) = 0 \end{cases}
\end{equation}

\subsection{Betweenness Centrality}
The metrics used were \emph{in-strength} and \emph{betweenness centrality}\cite{Brandes:2001p2678}. 
Centrality measurements can provide safe indicators of the importance of a node solely based on the topology of the network. The \emph{betweenness centrality} measures the importance of vertices by taking into account the number of shortest paths that pass through each vertex in a network. The betweenness centrality, $C_B(i)$, of vertex $i$ is defined as the sum of ratios of the number of total shortest paths that pass through $i$, $\sigma_{st}(i)$, by the total count of shortest paths, $\bar{\sigma}_{st}$, considering every pair of vertices, $(s, t)$, as given by:
\begin{equation}
\label{eq:betweenness_centrality} 
C_B(i) = \sum_{\underbrace{s, t}_{s\neq t \neq i}}{\sigma_{st}(i)\over \bar{\sigma}_{st}}
\end{equation}

\begin{figure*}[p]
  \centering
  \includegraphics[width=11cm]{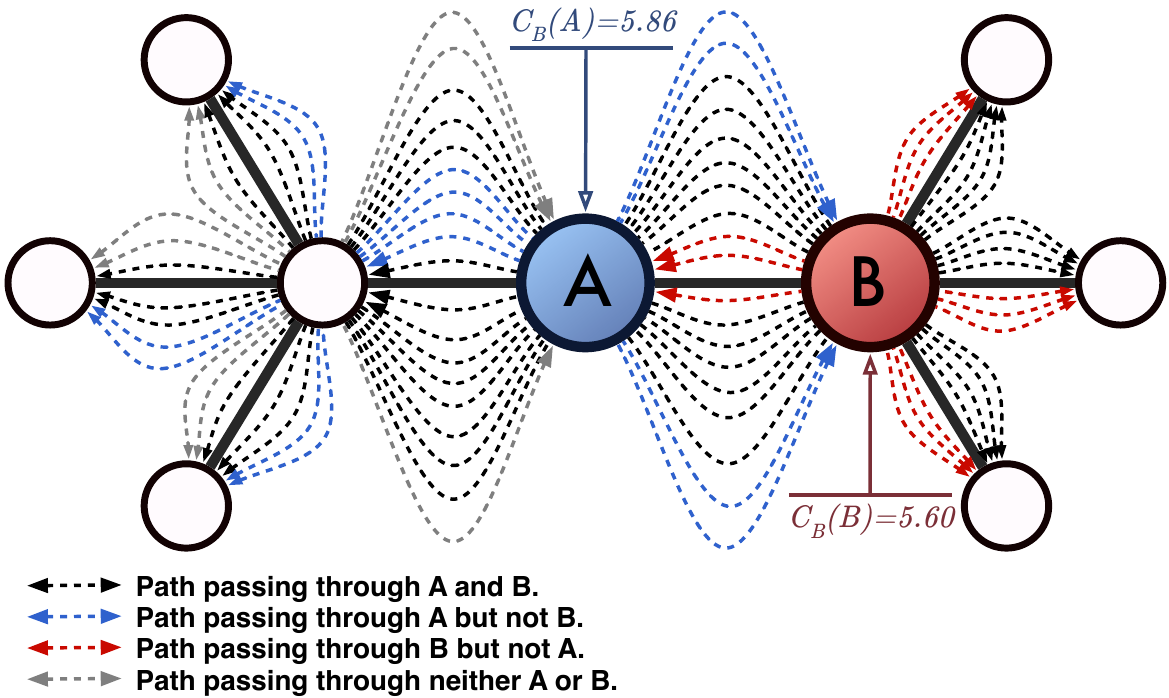}
 \caption{{\bf Illustrative example of betweenness centrality.} Network with 2 highlighted nodes, $A$ and $B$ presenting their respective betweenness centralities, $C_B(1)$ and $C_B(2)$, obtained with equation~\ref{eq:betweenness_centrality}. The paths between each pair of nodes are shown in different colors according to the legends. Node $A$ has only $2$ connections while $B$ has $4$ connections. However, $A$ is much more central than $B$.
}
  \label{fig:betweennessExample}
\end{figure*}

Unlike the traditional node degree, centrality measurements take into account all the vertices of the network resulting in a global overview of the network structure as seen in the example in figure~\ref{fig:betweennessExample}.

Classification of scientific papers and journals by subject is one of the most difficult and yet essential problems of information science. While the JCR subject categories are indicative of the main fields of a journal, they may fail to describe its interdisciplinarity because of the low diversity of subject categories for each node, barely surpassing $2$ subject categories per node. 

Much richer information about interdisciplinarity can be obtained by considering not only the individual categories of a journal, but rather the subject categories of journals that frequently cite it. For example, a journal with subject category of \emph{physics} bringing contributions in \emph{biological physics} is likely to be cited by journals classified as \emph{physics} and \emph{biology}. Such information can be obtained directly in terms of the topology of the journals networks described in the previous section.

Considering the first neighborhood of in-edges for each node in the journals citation network, i.e. journals that cite a journal representing the node under analysis, one can count the frequency of appearance of each subject category for such nodes, as shown in Figure~\ref{fig:methodology}. As a result, every node can be coupled with a histogram that provides information about the related subject categories of a journal, as well of its importance. Thus journals can also be reclassified according to the subjects appearing in its citation neighborhood.

\subsection{Network Visualization}
Network visualization methodologies may provide interesting insights about the correspondence of features and topological structure of networks. Traditionally, complex networks are visualized by placing nodes as geometric shapes over a plane or 3D space, while edges are represented by lines connecting them. Choosing the projected positions of nodes is one of the major challenges of this methodology and can be addressed in various ways\cite{Davidson:1996uq,Cohen:1997kx,Chen:2009ly}. Force-directed methods\cite{KAMADA:1989lr,FRUCHTERMAN:1991fk} provide a general way to place nodes in any metric space and can be applied to a wide range of networks, with which visually appealing results may be obtained. They work by initially placing the nodes over a metric space at random positions, then obtaining the configuration of minimal potential energy of a system as if each node was interacting by physical forces.

Here we employed the \emph{Fruchterman-Reingold} algorithm (FR)\cite{FRUCHTERMAN:1991fk}, which is a force-directed method using both attractive and repulsive forces in order to place the nodes of a network over a 2D or 3D space. A pair of nodes interact by repulsive Coulomb-based forces, $F_{(r)j}$. Nodes connected by edges($(i,j)\,\in\,\mathcal{E}$) also interact by attractive squared version of the Hook law force, $F_{(a)j}$, as described in equation~\ref{eq_force_electric}.

\begin{eqnarray} 
	\vec{F}_{(a)j} &=&\sum_{(i,j)} a (\vec{R}_i-\vec{R}_j)^2 \hat{r}_{ij}\\
	\vec{F}_{(r)j} &=&\sum_{i\,\in\,\mathcal{V}} {-b_{ij}\over (\vec{R}_i-\vec{R}_j)^2}\hat{r}_{ij}
 \label{eq_force_electric}
\end{eqnarray}

By minimizing the energy of this linear system, one should obtain a set of positions for each vertex in a way that the preferred Euclidian distance between each connected pair is obtained from equation~\ref{eq:prefferedDistance}. 

\begin{equation}
\label{eq:prefferedDistance}
d_{ij}^{*}=\left({b\over a}\right)^{1\over4}
\end{equation}

This methodology can be extended to edge weighted networks by simply making the attractive force constant, $a$, dependent on the edge weight, $w_{ij}$. Therefore, $a_{ij} = a w_{ij}^{4}$ so that $d_{ij}^{*} \propto  w_{ij}^{-1}$.

Solving the system of differential equations with the complete set of repulsive interactions between pairs of nodes is a n-body problem. Further optimizations such as the Fast Multipole Method\cite{Greengard:1987p908} can be applied to make this methodology computationally viable.

\end{document}